\documentclass[11pt,leqno]{article}
\usepackage{a4}
\usepackage[T1]{fontenc}
\usepackage[english]{babel}
\usepackage{latexsym}
\usepackage{amsmath}
\usepackage{amssymb}
\usepackage{mathrsfs}
\usepackage{color}
\usepackage{cite}
\usepackage{titling}
\usepackage{bbold}
\makeatletter
\@addtoreset{equation}{section} 
\makeatother
\definecolor{move}{rgb}{.3,.1,.8}
\newcommand{\ud}{\mathrm{d}}
\newcommand{\ds}{\displaystyle}

\newcommand{\e}{\mathrm{e}}

\newcommand{\R}{\mathbb{R}}
\newcommand{\N}{\mathbb{N}}
\newcommand{\Q}{\mathbb{Q}}

\newenvironment{pr}{\vspace{5pt}\textbf{{\small Proof :}}\\}{\hspace{\stretch{1}}\rule{1ex}{1ex}\vspace{5pt}}
\newtheorem{thm}{Theorem}[section]

\usepackage{fancyhdr}
\fancypagestyle{plain}{\fancyhead{}}
\title{Rapid exponential stabilization of a 1-D transmission wave equation with in-domain anti-damping}
\author{{FATHI HASSINE}\\ \textit{UR Analysis and Control of PDE UR13ES64}\\ \textit{Department of Mathematics, Faculty of Sciences of Monastir}\\ \textit{University of Monastir, 5019 Monastir, Tunisia}\\ \textit{email:} \texttt{fathi.hassine@fsm.rnu.tn}}
\date{}
\begin{document}
\maketitle
\begin{center}
\abstract{We consider the problem of pointwise stabilization of a one-dimensional wave equation with an internal spatially varying anti-damping term. We design a feedback law based on the backstepping method and prove exponential stability of the closed-loop system with a desired decay rate.}
\end{center}
\section{Introduction}
Originally developed for spatially one-dimensional linear parabolic systems the so-called backstepping method has been generalized to a broader class of distributed parameter systems comprising particular higher-dimensional equations (see~\cite{M} and the contained references), nonlinear equations~\cite{VK1,CeCr,CeCo}, parabolic equations~\cite{L} and hyperbolic equations~\cite{SCK,SK2,SK3}. However, results concerning systems with more general actuation such as the pointwise control are very rare (see~\cite{WW,WWK}).

In recent years a lot of papers were devoted to the study of elastic structures with pointwise actuator. Surprisingly, the control properties of those systems are very different depending on where the controls are located and which kind of boundary conditions are applied (see~\cite{AHT,AJ,AJK,CCW,Ha3,Ho,JTZ,Tuc} for the wave equation,~\cite{AT,C,R} for the beam equation and~\cite{AJM,AM} for the wave/beam equation) where the authors give several examples (or conditions on the actuator location) showing that both uniform and nonuniform decay (strong stability or non-strong stability) (controllability or uncontrollability properties) may occur using essentially the frequency domains method and the multiplier techniques for problem of stabilization and Hilbert uniqueness method for problems of control. All the above-tools can be reformulated in semigroups theory. Strongly continuous semigroups remain an efficient tool for investigating the stability or to build a control for unstable systems.

In this paper, we consider the problem of pointwise stabilization of a one-dimensional wave equation with an internal anti-damping term with Dirichlet boundary conditions in left and right hand side of a unit interval. Without the anti-damping term, Ho~\cite{Ho} shows that a such system is approximately controllable (in some time $T>0$) if and only if the point $\xi$ where the feedback is localized is an irrational number. However, he shows also that even when $\xi$ is not a rational number the system is not exactly controllable. Noting that the situation would be different if we change the boundary condition at one of the endpoints to a Newman homogenous boundary condition and keeping the boundary condition at the other end the same. Indeed, under some geometrical conditions on $\xi$ the system system remains approximately controllable and for more restrictive conditions on $\xi$ the system remains approximately controllable.

In this perspective, in this paper we would like to design a control that does not depend to the position of a such actuator and makes the system rapidly stable. From the engineering point of view this will be not only useful but also so practical, first, because in practical it is ''impossible'' to localized for instance a point $\xi$ as an irrational number on a string! Therefore, because our control design is based on the method of backstepping~\cite{SK1,SK2,KGBS}, the gain functions formula is explicit and can be calculated numerically via a scheme of successive approximation. This makes its implementation possible in real problems. In comparison with the existing literature the novelty of this paper is the explicit construction of the feedback laws and the complete solving of the strange coupled boundary value problem mentioned above even with the actuation of the anti-damping term which makes the system anti-stable in the sense that the eigenvalues of the open-loop system can all be in the right half of the complex plane, which produces an exponential growth of the norm of the solutions.

The paper is organized as follows. In section~\ref{ctw30} we formulate the problem and state the main result. In section~\ref{ctw31} we introduce the transformation and the boundary feedback which transform the plant into the target system. In section~\ref{ctw32}, we show that this target PDE is exponentially stable.
\section{Preliminaries}\label{ctw30}
We consider the following transmission wave equation
\begin{equation}\label{ctw1}
\left\{\begin{array}{ll}
\ddot{u}_{1}(x,t)=u_{1}''(x,t)+2\lambda_{1}(x)\dot{u}_{1}(x,t)+\alpha_{1}(x)u_{1}'(x,t)+\beta_{1}(x)u_{1}(x,t)&\text{in }(0,\xi)\times(0,+\infty)
\\
\ddot{u}_{2}(x,t)=u_{2}''(x,t)+2\lambda_{2}(x)\dot{u}_{2}(x,t)+\alpha_{2}(x)u_{2}'(x,t)+\beta_{2}(x)u_{2}(x,t)&\text{in }(\xi,1)\times(0,+\infty)
\\
u_{1}(\xi,t)=u_{2}(\xi,t)&\text{for } t\in (0,+\infty)
\\
u_{1}'(\xi,t)=u_{2}'(\xi,t)+U(t)&\text{for } t\in (0,+\infty)
\\
u_{1}(0,t)=u_{2}(1,t)=0&\text{for } t\in (0,+\infty)
\\
u_{1}(x,0)=u_{1}^{0}(x),\;\dot{u}_{1}(x,0)=u_{1}^{1}(x)&\text{for }x\in(0,\xi)
\\
u_{2}(x,0)=u_{1}^{0}(x),\;\dot{u}_{2}(x,0)=u_{2}^{1}(x)&\text{for }x\in(\xi,1),
\end{array}\right.
\end{equation}
where the dot stands for the time derivative, the prime stands for the space derivative and for each time $t\geq 0$ the states of the system $u_{1}(\,.\,,t)$ and $u_{2}(\,.\,,t)$ and the input $U(t)$ are real-valued functions. The functions $u_{1}^{0}$, $u_{1}^{1}$, $u_{2}^{0}$ and $u_{2}^{1}$ are the initial conditions and the function $\alpha_{1}$, $\beta_{1}$, $\lambda_{1}$, $\alpha_{2}$, $\beta_{2}$, $\lambda_{2}$ are the coefficients whose regularity will be defined later, in particular $\lambda_{1}$ and $\lambda_{2}$ are the anti-damping terms. 

It is well know that the open-loop plant (i.e., with $U(t)=0$) is unstable and all the eigenvalues of the system are located in the right half of the complex plan when $\lambda_{1}$ and $\lambda_{2}$ are positive and $\alpha_{1}=\beta_{1}=\alpha_{2}=\beta_{2}=0$.

When the feedback is given by $U(t)=\dot{u}_{1}(\xi,t)$ and all the coefficients are null (i.e., $\alpha_{1}=\beta_{1}=\lambda_{1}=\alpha_{2}=\beta_{2}=\lambda_{2}=0$) the strong stability of energy for the model~\eqref{ctw1} is provided if and only if $\xi$ is an irrational number (see~\cite{CCW} and~\cite{Ho}). Furthermore, for any $\xi\in(0,1)\backslash\Q$ the decay of the solution is not uniform in the energy space. For non symmetric boundary conditions (i.e., Dirichlet boundary condition on one side and Newman boundary condition on the other side) the uniform exponential stability holds if and only if $\ds \xi=\frac{p}{q}$ with $p$ is odd (see~\cite{Ho}) and where the fastest decay rate of the solution is obtained when the actuator is located at the middle of the string (see~\cite{AHT,AN}). Besides, if $\xi$ satisfies a Diophantine approximations properties then we have polynomial decay rate for the regular data (see~\cite{Tuc},~\cite{AHT} and~\cite{JTZ}). In our case of symmetric boundary conditions (Dirichlet boundary condition on both sides) Tucsnak~\cite{Tuc} proved that for every $\xi\in(0,1)\backslash\Q$ there exists $\ds\psi_{\xi}:[0,+\infty)\longrightarrow\R$ with $\ds\lim_{t\longrightarrow+\infty}\psi_{\xi}(t)=0$ such that the solution $(u_{1},u_{2})$ of~\eqref{ctw1} decreases with the rate that of $\psi_{\xi}(t)$ as $t$ goes to the infinity for regular data, where the function $\psi_{\xi}$ tends to zero at most as $\ds\frac{1}{t}$. Recently, it was proved in~\cite{Ha3} that $\psi_{\xi}$ can be of the form $\ds\frac{1}{\ln^{2}(1+t)}$ and this for all $\xi$ satisfying some geometrical condition.

Noting that without lost of generality, we set $\alpha_{1}\equiv0$ and $\alpha_{2}\equiv0$. Indeed, if $\alpha_{1}$ or  $\alpha_{2}$ is not identically zero, the following rescaling of the state variables,
$$
\tilde{u}_{1}(x,t)=\e^{\frac{1}{2}\int_{\xi}^{x}\alpha_{1}(\tau)\,\ud\tau}u_{1}(x,t)\quad\text{and}\quad\tilde{u}_{2}(x,t)=\e^{\frac{1}{2}\int_{\xi}^{x}\alpha_{2}(\tau)\,\ud\tau}u_{2}(x,t),
$$
would transform the original wave equation into another one that does not have the first-order spatial derivative terms, namely we obtain the following system
\begin{equation}\label{ctw2}
\left\{\begin{array}{ll}
\ddot{u}_{1}(x,t)=u_{1}''(x,t)+2\lambda_{1}(x)\dot{u}_{1}(x,t)+\beta_{1}(x)u_{1}(x,t)&\text{in }(0,\xi)\times(0,+\infty)
\\
\ddot{u}_{2}(x,t)=u_{2}''(x,t)+2\lambda_{2}(x)\dot{u}_{2}(x,t)+\beta_{2}(x)u_{2}(x,t)&\text{in }(\xi,1)\times(0,+\infty)
\\
u_{1}(\xi,t)=u_{2}(\xi,t)&\text{for } t\in (0,+\infty)
\\
u_{1}'(\xi,t)=u_{2}'(\xi,t)+\frac{1}{2}(\alpha_{1}(\xi)-\alpha_{2}(\xi))u_{1}(\xi,t)+U(t)&\text{for } t\in (0,+\infty)
\\
u_{1}(0,t)=u_{2}(1,t)=0&\text{for } t\in (0,+\infty)
\\
u_{1}(x,0)=u_{1}^{0}(x),\;\dot{u}_{1}(x,0)=u_{1}^{1}(x)&\text{for }x\in(0,\xi)
\\
u_{2}(x,0)=u_{1}^{0}(x),\;\dot{u}_{2}(x,0)=u_{2}^{1}(x)&\text{for }x\in(\xi,1),
\end{array}\right.
\end{equation}
where for the sake of simplicity and clarity we keep the same notations for the systems~\eqref{ctw1} and~\eqref{ctw2} even that the coefficients of both systems are different. Noting that all our analysis will be done accordingly to the system~\eqref{ctw2}.

The main idea of this paper is to use the transformations
\begin{equation}\label{ctw3}
w_{1}(x,t)=h_{1}(x)u_{1}(x,t)-\int_{0}^{x}k_{1}(x,y)u_{1}(y,t)\,\ud y-\int_{0}^{x}s_{1}(x,y)\dot{u}_{1}(y,t)\,\ud y,\qquad0<x<\xi
\end{equation}
and
\begin{equation}\label{ctw4}
w_{2}(x,t)=h_{2}(x)u_{2}(x,t)-\int_{1}^{x}k_{2}(x,y)u_{2}(y,t)\,\ud y-\int_{1}^{x}s_{2}(x,y)\dot{u}_{2}(y,t)\,\ud y,\qquad\xi<x<1
\end{equation}
and the feedback
\begin{equation}\label{ctw5}
\begin{split}
U(t)&=\frac{u_{1}(\xi,t)}{h_{1}(\xi)}\left(h_{2}'(\xi)-h_{1}'(\xi)+k_{1}(\xi,\xi)-k_{2}(\xi,\xi)\right)+\frac{u_{1}(\xi,t)}{2}\left(\alpha_{2}(\xi)-\alpha_{1}(\xi)\right)
\\
&+\frac{1}{h_{1}(\xi)}\bigg(\int_{0}^{\xi}k_{1x}(\xi,y)u_{1}(y,t)\,\ud y-\int_{1}^{\xi}k_{2x}(\xi,y)u_{2}(y,t)\,\ud y+\int_{0}^{\xi}s_{1x}(\xi,y)\dot{u}_{1}(y,t)\,\ud y
\\
&-\int_{1}^{\xi}s_{2x}(\xi,y)\dot{u}_{2}(y,t)\,\ud y\bigg)+\left(s_{1}(\xi,\xi)-s_{2}(\xi,\xi)\right)\frac{\dot{u}_{1}(\xi,t)}{h_{1}(\xi)}.
\end{split}
\end{equation}
where $h_{1}$, $h_{2}$ and the kernels $k_{1}$, $k_{2}$, $s_{1}$ and $s_{2}$ are appropriately chosen to convert the original system~\eqref{ctw2} into the following one
\begin{equation}\label{ctw6}
\left\{\begin{array}{ll}
\ddot{w}_{1}(x,t)=w_{1}''(x,t)-2d_{1}(x)\dot{w}_{1}(x,t)-c_{1}(x)w_{1}(x,t)&\text{in }(0,\xi)\times(0,+\infty)
\\
\ddot{w}_{2}(x,t)=w_{2}''(x,t)-2d_{2}(x)\dot{w}_{2}(x,t)-c_{2}(x)w_{2}(x,t)&\text{in }(\xi,1)\times(0,+\infty)
\\
w_{1}(\xi,t)=w_{2}(\xi,t)&\text{for } t\in (0,+\infty)
\\
w_{1}'(\xi,t)=w_{2}'(\xi,t)&\text{for } t\in (0,+\infty)
\\
w_{1}(0,t)=w_{2}(1,t)=0&\text{for } t\in (0,+\infty)
\\
w_{1}(x,0)=w_{1}^{0}(x),\;\dot{w}_{1}(x,0)=w_{1}^{1}(x)&\text{for }x\in(0,\xi)
\\
w_{2}(x,0)=w_{1}^{0}(x),\;\dot{w}_{2}(x,0)=w_{2}^{1}(x)&\text{for }x\in(\xi,1),
\end{array}\right.
\end{equation}
with appropriate function $d_{1},\,d_{2},\,c_{1}$ and $c_{2}$ so that this new system is exponentially stable. The functions $d_{1},\,d_{2},\,c_{1}$ and $c_{2}$  can always be chosen to provide any desired decay rate.

Then, we use the exponential stability of~\eqref{ctw6} and the irreversibility of the transformations~\eqref{ctw3} and~\eqref{ctw4} to obtain stability of the closed-loop system~\eqref{ctw2} with the feedback~\eqref{ctw5}.

For $s=1,2$ introducing the spaces $H_{L}^{s}(0,\xi)$ and $H_{R}^{s}(\xi,1)$ by
$$
H_{L}^{s}(0,\xi)=\{z\in H^{s}(0,\xi):\;z(0)=0\}\text{ and }H_{R}^{s}(\xi,1)=\{z\in H^{s}(\xi,1):\;z(1)=0\}
$$
endowed with the $H^{s}$-norm and the domains
$$
\mathcal{T}_{1}=\{(x,y)\in\R^{2}:\;0\leq x\leq\xi,\;0\leq y\leq x\}\text{ and }\mathcal{T}_{2}=\{(x,y)\in\R^{2}:\;\xi\leq x\leq 1,\;x\leq y\leq 1\}.
$$
\begin{thm}\label{ctw51}
We suppose that $\lambda_{1}\in\mathrm{C}^{2}([0,\xi])$, $\lambda_{2}\in\mathrm{C}^{2}([\xi,1])$, $\beta_{1}\in\mathrm{C}^{1}([0,\xi])$ and $\beta_{2}\in\mathrm{C}^{1}([\xi,1])$ then there exist $h_{1}\in\mathrm{C}^{2}([0,\xi])$, $h_{2}\in\mathrm{C}^{2}([\xi,1])$, $k_{1},\,s_{1}\in\mathrm{C}^{2}(\mathcal{T}_{1})$ and $k_{2},\,s_{2}\in\mathrm{C}^{2}(\mathcal{T}_{2})$ such that for any $(u_{1}^{0},u_{2}^{0},u_{1}^{1},u_{2}^{1})\in H^{1}_{L}(0,\xi)\times H_{R}^{1}(\xi,1)\times L^{2}(0,\xi)\times L^{2}(\xi,1)$ satisfying the following compatibility conditions
\begin{equation}\label{ctw48}
\begin{split}
h_{1}(\xi)u_{1}^{0}(\xi)-\int_{0}^{\xi}k_{1}(\xi,y)u_{1}^{0}(y)\,\ud y-\int_{0}^{\xi}s_{1}(\xi,y)u_{1}^{1}(y)\,\ud y=
\\
h_{2}(\xi)u_{2}^{0}(\xi)-\int_{1}^{\xi}k_{2}(\xi,y)u_{2}^{0}(y)\,\ud y-\int_{1}^{\xi}s_{2}(\xi,y)u_{2}^{1}(y)\,\ud y
\end{split}
\end{equation}
and
\begin{equation}\label{ctw49}
\begin{split}
h_{1}(\xi)(u_{1}^{0})'(\xi)+h_{1}'(\xi)u_{1}^{0}(\xi)-k_{1}(\xi,\xi)u_{1}^{0}(\xi)-s_{1}(\xi,\xi)u_{1}^{1}(\xi)-\int_{0}^{\xi}k_{1x}(\xi,y)u_{1}^{0}(y)\,\ud y
\\
-\int_{0}^{\xi}s_{1x}(\xi,y)u_{1}^{1}(y)\,\ud y=h_{2}(\xi)(u_{2}^{0})'(\xi)+h_{2}'(\xi)u_{2}^{0}(\xi)-k_{2}(\xi,\xi)u_{2}^{0}(\xi)-s_{2}(\xi,\xi)u_{2}^{1}(\xi)
\\
-\int_{1}^{\xi}k_{2x}(\xi,y)u_{2}^{0}(y)\,\ud y-\int_{1}^{\xi}s_{2x}(\xi,y)u_{2}^{1}(y)\,\ud y,
\end{split}
\end{equation}
such that the system~\eqref{ctw2} with the feedback law~\eqref{ctw5} has a unique solution in the space $\mathrm{C}([0,+\infty),H^{1}_{L}(0,\xi)\times H^{1}_{R}(\xi,1))\cap\mathrm{C}^{1}([0,+\infty),L^{2}(0,\xi)\times L^{2}(\xi,1))$. Moreover, for any $\omega>0$, there exists constant $C>0$ independent of the initial data such that the solution satisfy
\begin{equation}\label{ctw7}
\begin{split}
\|(u_{1}(\,.\,,t),u_{2}(\,.\,,t),\dot{u}_{1}(\,.\,,t),\dot{u}_{2}(\,.\,,t))\|_{H^{1}(0,\xi)\times H^{1}(\xi,1)\times L^{2}(0,\xi)\times L^{2}(\xi,1)}
\\
\leq C\e^{-\omega t}\|(u_{1}^{0},u_{2}^{0},u_{1}^{1},u_{2}^{1})\|_{H^{1}(0,\xi)\times H^{1}(\xi,1)\times L^{2}(0,\xi)\times L^{2}(\xi,1)}.
\end{split}
\end{equation}
\end{thm}

Noting here that the proof of Theorem~\ref{ctw43} provides a numeric computation scheme of successive approximation to compute the kernel functions $k_{1},\,k_{2},\,s_{1}$ and $s_{2}$ in our feedback laws~\eqref{ctw5}. This makes the feedback laws~\eqref{ctw5} implementable in real problems.
\section{Control design}\label{ctw31}
In this section we derive the equations for the functions $h_{1}$, $h_{2}$, $k_{1}$, $k_{2}$, $s_{1}$ and $s_{2}$ and show that they have a unique twice continuously differentiable solution. Noting that in this section we will omit all the subscripts with the numbers $1$ and $2$ and we will recall them only when it is necessary. For instance, when we talk about $w$ this mean that $w$ play the role for both $w_{1}$ and $w_{2}$. We denote also by $\varepsilon_{1}=0$ and $\varepsilon_{2}=1$.  

Using the transformations~\eqref{ctw3} and~\eqref{ctw4} and equation~\eqref{ctw3}, we get
\begin{equation*}
\begin{split}
\ddot{w}-w''+2d(x)\dot{w}+c(x)w=\int_{\varepsilon}^{x}u(y)\Big[k_{xx}-k_{yy}-(c(x)+\beta(y))k-2(\lambda(y)+d(x))s_{yy}
\\
-2(\lambda(y)\beta(y)+\lambda''(y)+d(x)\beta(y))s-4\lambda'(y)s_{y}\Big]\,\ud y
\\
+\int_{\varepsilon}^{x}\dot{u}(y)\Big[s_{xx}-s_{yy}-2(\lambda(y)+d(x))k-(4\lambda^{2}(y)+4d(x)\lambda(y)+c(x)+\beta(y))s\big]\,\ud y
\\
+u(x)\Big[2k'(x,x)+2(\lambda(x)+d(x))s_{y}(x,x)+2\lambda'(x)s(x,x)+(c(x)+\beta(x))h(x)-h''(x)\Big]
\\
+2\dot{u}(x)\Big[s'(x,x)+(\lambda(x)+d(x))h(x)\Big]-2u'(x)\Big[(\lambda(x)+d(x))s(x,x)+h'(x)\Big]
\\
+u'(\varepsilon)\Big[k(x,\varepsilon)+2(\lambda(\varepsilon)+d(x))s(x,\varepsilon)\Big]+s(x,\varepsilon)\dot{u}'(\varepsilon).
\end{split}
\end{equation*}
In order to satisfy~\eqref{ctw6} we choose $k$ and $s$ as solution of
\begin{eqnarray}
k_{xx}(x,y)-k_{yy}(x,y)&\!\!\!=\!\!\!&(c(x)+\beta(y))k(x,y)+2(\lambda(y)+d(x))s_{yy}(x,y)\label{ctw8}
\\
&&+2(\lambda(y)\beta(y)+\lambda''(y)+d(x)\beta(y))s(x,y)+4\lambda'(y)s_{y}(x,y)\nonumber
\\
2k'(x,x)&\!\!\!=\!\!\!&-2(\lambda(x)+d(x))s_{y}(x,x)-2\lambda'(x)s(x,x)\label{ctw9}
\\
&&-(c(x)+\beta(x))h(x)+h''(x)\nonumber
\\
k(x,\varepsilon)&\!\!\!=\!\!\!&0,\label{ctw10}
\end{eqnarray}
and
\begin{eqnarray}
s_{xx}(x,y)-s_{yy}(x,y)&\!\!\!=\!\!\!&2(\lambda(y)+d(x))k(x,y)\label{ctw11}
\\
&&+(4\lambda^{2}(y)+4d(x)\lambda(y)+c(x)+\beta(y))s(x,y)\nonumber
\\
s'(x,x)&\!\!\!=\!\!\!&-(\lambda(x)+d(x))h(x)\label{ctw12}
\\
(\lambda(x)+d(x))s(x,x)&\!\!\!=\!\!\!&-h'(x)\label{ctw13}
\\
s(x,\varepsilon)&\!\!\!=\!\!\!&0.\label{ctw14}
\end{eqnarray}
The following result asserting the existence of the kernel functions $h_{1},\,h_{2},\,k_{1},\,k_{2},\,s_{1}$ and $s_{2}$ satisfying the equations~\eqref{ctw8}-\eqref{ctw14}.
\begin{thm}\label{ctw43}
Let $d_{1},\,c_{1}\in\mathrm{C}^{1}([0,\xi])$ and $d_{2},\,c_{2}\in\mathrm{C}^{1}([\xi,1])$. Then~\eqref{ctw8}-\eqref{ctw14} have a unique solution $h_{1}\in\mathrm{C}^{2}([0,\xi])$, $h_{2}\in\mathrm{C}^{2}([\xi,1])$, $k_{1},\,s_{1}\in\mathrm{C}^{2}(\mathcal{T}_{1})$ and $k_{2},\,s_{2}\in\mathrm{C}^{2}(\mathcal{T}_{2})$ such that $h_{1}(0)=1$ and $h_{1}(\xi)=h_{2}(\xi)$.
\end{thm}
\begin{pr}
Dividing \eqref{ctw12} by \eqref{ctw13}, we have $s'(x,x)s(x,x)=h'(x)h(x)$ then integrating we get $h(x)^{2}=s(x,x)^{2}+A$. From~\eqref{ctw14} we find that $A=h(\varepsilon)^{2}$. Using~\eqref{ctw13} we obtain
$$
\frac{h'(x)}{\sqrt{h(x)^{2}-A}}=\lambda(x)+d(x)
$$
which gives
\begin{equation}\label{ctw15}
h(x)=\sqrt{A}\cosh\left(\int_{\varepsilon}^{x}a(\tau)\,\ud\tau\right)
\end{equation}
where
\begin{equation}\label{ctw16}
a(x)=\lambda(x)+d(x).
\end{equation}
Since $h_{1}(0)=1$ and $h_{1}(\xi)=h_{2}(\xi)$ so that $A_{1}=1$ and $\ds A_{2}=\frac{\cosh^{2}\left(\ds\int_{0}^{\xi}a(\tau)\,\ud\tau\right)}{\cosh^{2}\left(\ds\int_{1}^{\xi}a(\tau)\,\ud\tau\right)}$. Thus, we can write
\begin{equation*}
s(x,x)=-\frac{h'(x)}{a(x)}=-\sqrt{A}\sinh\left(\int_{\varepsilon}^{x}a(\tau)\,\ud\tau\right).
\end{equation*}

Our goal is to find $k(x,x)$ explicitly. Let us denote $f(x)=s_{y}(x,x)$, integrating~\eqref{ctw9} and using~\eqref{ctw10}, we obtain
\begin{equation}\label{ctw17}
2k(x,x)=h'(x)+\int_{\varepsilon}^{x}\left[-2a(\tau)f(\tau)+2\frac{\lambda'(\tau)h'(\tau)}{a(\tau)}-(\beta(\tau)+c(\tau))h(\tau)\right]\ud\tau.
\end{equation}
We see that we have to find $f(x)$ in order to get $k(x,x)$. From~\eqref{ctw12} we get
$$
s_{x}(x,x)=-a(x)h(x)-f(x).
$$
which gives 
$$
s_{xx}(x,x)-s_{yy}(x,x)=(s_{x}(x,x)-s_{y}(x,x))'=[-a(x)h(x)-2f(x)]'.
$$
Using~\eqref{ctw12},~\eqref{ctw16} and~\eqref{ctw17} and the previous equation we show that $f$ is the solution of the integro-differential equation
\begin{equation}\label{ctw18}
\left\{\begin{array}{l}
\ds 2f'(x)-2a(x)\int_{\varepsilon}^{x}a(\tau)\ud\tau=L(x)
\\
f(\varepsilon)=-\sqrt{A}a(\varepsilon),
\end{array}\right.
\end{equation}
where $L$ is defined by
\begin{equation}\label{ctw19}
\begin{split}
L(x)=4\lambda(x)h'(x)-(c(x)+\beta(x))\frac{h'(x)}{a(x)}-2a(x)h'(x)-a'(x)h(x)
\\
-2a(x)\int_{\varepsilon}^{x}\frac{\lambda'(\tau)h'(\tau)}{a(\tau)}\ud\tau+a(x)\int_{\varepsilon}^{x}(\beta(\tau)+c(\tau))h(\tau)\ud\tau.
\end{split}
\end{equation}
From~\eqref{ctw18} we obtain the following second order ordinary differential equation
\begin{equation}\label{ctw20}
\left\{\begin{array}{l}
2a(x)f''(x)-2a'(x)f'(x)-2a(x)^{3}f(x)=L'(x)a(x)-L(x)a'(x)
\\
\ds f(\varepsilon)=-\sqrt{A}a(\varepsilon),\qquad f'(\varepsilon)=-\frac{\sqrt{A}}{2}a'(\varepsilon).
\end{array}\right.
\end{equation}
The solution of~\eqref{ctw20} is
\begin{equation}\label{ctw21}
f(x)=-\sqrt{A}a(\varepsilon)\cosh\left(\int_{\varepsilon}^{x}a(\tau)\ud\tau\right)+\frac{1}{2}\int_{\varepsilon}^{x}L(y)\cosh\left(\int_{y}^{x}a(\tau)\,\ud\tau\right)\,\ud y.
\end{equation}
It is easy to show that
\begin{equation}\label{ctw33}
\sqrt{A}a(\varepsilon)\int_{\varepsilon}^{x}a(\tau)\cosh\left(\int_{\varepsilon}^{\tau}a(t)\,\ud t\right)\ud\tau=a(\varepsilon)\frac{h'(x)}{a(x)}.
\end{equation}
Using the fact that, $\ds A\cosh\left(\int_{y}^{\tau}a(t)\,\ud t\right)=h(\tau)h(y)-\frac{h'(\tau)}{a(\tau)}\frac{h'(y)}{a(y)}$ then integrating and using the fact that $\ds\left(\frac{h'(x)}{a(x)}\right)'=a(x)h(x)$ then it follows
\begin{equation}\label{ctw36}
\begin{split}
-\frac{\sqrt{A}}{2}\int_{\varepsilon}^{x}a(\tau)\int_{\varepsilon}^{\tau}a(y)^{2}\sinh\left(\int_{\varepsilon}^{y}a(t)\,\ud t\right)\cosh\left(\int_{y}^{\tau}a(t)\,\ud t\right)\,\ud y\,\ud\tau=
\\
\frac{h'(x)h(x)^{2}}{2A}-\frac{h'(x)}{2Aa(x)}\left(\int_{\varepsilon}^{x}a'(y)h(y)^{2}\,\ud y+A.a(\varepsilon)\right)-\frac{h(x)}{A}\int_{\varepsilon}^{x}h'(y)^{2}\,\ud y.
\end{split}
\end{equation}
and
\begin{equation}\label{ctw37}
\begin{split}
-\frac{1}{2}\int_{\varepsilon}^{x}a(\tau)\int_{\varepsilon}^{\tau}a'(y)h(y)\cosh\left(\int_{y}^{\tau}a(t)\,\ud t\right)\,\ud y\,\ud\tau=
\\
\frac{h'(x)}{2Aa(x)}\int_{\varepsilon}^{x}a'(y)h^{2}(y)\,\ud y-\frac{h(x)}{2A}\int_{\varepsilon}^{x}\frac{h'(y)h(y)a'(y)}{a(y)}\,\ud y.
\end{split}
\end{equation}
Since by integration by parts we have
$$
\frac{h(x)}{2A}\int_{\varepsilon}^{x}\frac{h'(y)h(y)a'(y)}{a(y)}\,\ud y=\frac{h'(x)h(x)^{2}}{2A}-\frac{h(x)}{2A}\int_{\varepsilon}^{x}h'(y)^{2}\,\ud y-\frac{h(x)}{2A}\int_{\varepsilon}^{x}a(y)^{2}h(y)^{2}\,\ud y,
$$
then the sum of~\eqref{ctw33},~\eqref{ctw36} and~\eqref{ctw37} gives
\begin{equation}\label{ctw40}
a(\varepsilon)\frac{h'(x)}{2a(x)}+\frac{h(x)}{2}\int_{\varepsilon}^{x}a(y)^{2}\,\ud y.
\end{equation}
Using the same argument as in~\eqref{ctw38} and~\eqref{ctw39} and the fact that $\ds h(x)^{2}+\left(\frac{h'(x)}{a(x)}\right)^{2}=A$ then by integration by parts, one gets
\begin{equation}\label{ctw34}
\begin{split}
-2\sqrt{A}.\int_{\varepsilon}^{x}a(\tau)\int_{\varepsilon}^{\tau}\lambda(y)a(y)\sinh\left(\int_{\varepsilon}^{y}a(t)\,\ud t\right)\cosh\left(\int_{y}^{\tau}a(t)\,\ud t\right)\,\ud y\,\ud\tau=
\\
\frac{h'(x)}{a(x)}\left(\lambda(\varepsilon)-\frac{\lambda(x)}{A}h(x)^{2}+\frac{1}{A}\int_{\varepsilon}^{x}\lambda'(y)h(y)^{2}\,\ud y\right)+\frac{2h(x)}{A}\int_{\varepsilon}^{x}\frac{\lambda(y)h'(y)^{2}}{a(y)}\,\ud y.
\end{split}
\end{equation}
and
\begin{equation}\label{ctw35}
\begin{split}
\int_{\varepsilon}^{x}a(\tau)\int_{\varepsilon}^{\tau}a(y)\int_{\varepsilon}^{y}\frac{\lambda'(t)h'(t)}{a(t)}\,\ud t\,\cosh\left(\int_{y}^{\tau}a(t)\,\ud t\right)\,\ud y\,\ud\tau=
\\
-\frac{h'(x)}{Aa(x)}\int_{\varepsilon}^{x}\frac{\lambda'(y)h'(y)^{2}}{a(y)^{2}}\,\ud y-\int_{\varepsilon}^{x}\frac{\lambda'(y)h'(y)}{a(y)}\,\ud y+\frac{h(x)}{A}\int_{\varepsilon}^{x}\frac{\lambda'(y)h'(y)h(y)}{a(y)}\,\ud y.
\end{split}
\end{equation}
The sum of~\eqref{ctw34} and~\eqref{ctw35} gives
\begin{equation}\label{ctw41}
\frac{h'(x)\lambda(x)}{a(x)}-h(x)\int_{\varepsilon}^{x}\lambda(y)a(y)\,\ud y-\int_{\varepsilon}^{x}\frac{\lambda'(y)h'(y)}{a(y)}\,\ud y.
\end{equation}
Using a simple integration by parts we can perform the following calculations
\begin{equation}\label{ctw38}
\begin{split}
-\frac{1}{2}\int_{\varepsilon}^{x}a(\tau)\int_{\varepsilon}^{\tau}a(y)\int_{0}^{y}(\beta(t)+c(t))h(t)\,\ud t\,\cosh\left(\int_{y}^{\tau}a(t)\,\ud t\right)\,\ud y\,\ud\tau=
\\
\frac{h'(x)}{2Aa(x)}\int_{\varepsilon}^{x}\frac{h(y)h'(y)}{a(y)}(\beta(y)+c(y))\,\ud y-\frac{h(x)}{2A}\int_{\varepsilon}^{x}h(y)^{2}(\beta(y)+c(y))\,\ud y
\\
+\frac{1}{2}\int_{\varepsilon}^{x}h(y)(\beta(y)+c(y))\,\ud y,
\end{split}
\end{equation}
and
\begin{equation}\label{ctw39}
\begin{split}
-\frac{1}{2}\int_{\varepsilon}^{x}a(\tau)\int_{\varepsilon}^{\tau}(\beta(y)+c(y))\cosh\left(\int_{y}^{\tau}a(t)\,\ud t\right)\sinh\left(\int_{\varepsilon}^{y}a(t)\,\ud t\right)\,\ud y\,\ud\tau=
\\
\frac{h(x)}{2A}\int_{\varepsilon}^{x}\left(\frac{h'(y)}{a(y)}\right)^{2}(\beta(y)+c(y))\,\ud y-\frac{h'(x)}{2Aa(x)}\int_{\varepsilon}^{x}\frac{h(y)h'(y)}{a(y)}(\beta(y)+c(y))\,\ud y.
\end{split}
\end{equation}
Summing~\eqref{ctw38} and~\eqref{ctw39} we obtain
\begin{equation}\label{ctw42}
\frac{1}{2}\int_{\varepsilon}^{x}(\beta(y)+c(y))h(y)\,\ud y-\frac{h(x)}{2}\int_{\varepsilon}^{x}(\beta(y)+c(y))\,\ud y.
\end{equation}
Putting~\eqref{ctw33},\eqref{ctw36},\eqref{ctw37},\eqref{ctw35},\eqref{ctw34},\eqref{ctw38} and \eqref{ctw39} into~\eqref{ctw21} where we have to recall~\eqref{ctw17} and~\eqref{ctw19} and using~\eqref{ctw40},\eqref{ctw41} and \eqref{ctw39} we find
\begin{equation}\label{ctw22}
\begin{split}
m(x):=k(x,x)&=\frac{h'(x)}{2a(x)}\left(2\lambda(x)+a(x)+a(\varepsilon)\right)
\\
&+\frac{h(x)}{2}\int_{\varepsilon}^{x}\left(d(y)^{2}-\lambda(y)^{2}-\beta(y)-c(y)\right)\,\ud y.
\end{split}
\end{equation}
Let us define $\rho_{i}(x,y)$ with i=1,\ldots,5 by
\begin{equation*}
\begin{array}{c}
\rho_{1}(x,y)=2(\lambda(y)+d(x)),\quad\rho_{2}(x,y)=c(x)+\beta(y),\quad\rho_{4}(x,y)=4\lambda'(y)
\\
\rho_{3}(x,y)=2(\lambda(y)\beta(y)+\lambda''(y)+d(x)\beta(y)),\quad\rho_{5}(x,y)=4\lambda^{2}(y)+4d(x)\lambda(y)+c(x)+\beta(y)
\end{array}
\end{equation*}
Then one get the following equation for the kernel functions
\begin{equation}\label{ctw23}
\left\{\begin{array}{rcl}
k_{xx}(x,y)-k_{yy}(x,y)&=&\rho_{1}(x,y)s_{yy}(x,y)+\rho_{2}(x,y)k(x,y)
\\
&&+\rho_{3}(x,y)s(x,y)+\rho_{4}(x,y)s_{y}(x,y)
\\
k(x,x)&=&m(x)
\\
k(x,0)&=&0,
\end{array}\right.
\end{equation}
and
\begin{equation}\label{ctw24}
\left\{\begin{array}{rcl}
s_{xx}(x,y)-s_{yy}(x,y)&=&\rho_{1}(x,y)k(x,y)+\rho_{5}(x,y)s(x,y)
\\
s(x,x)&=&m^{s}(x):=\ds-\sqrt{A}\sinh\left(\int_{\varepsilon}^{x}a(\tau)\,\ud\tau\right)
\\
s(x,0)&=&0.
\end{array}\right.
\end{equation}

To prove the existence of solutions of~\eqref{ctw23} and~\eqref{ctw24}, we perform the following change of variable
$$
\zeta=x+y,\qquad\eta=x-y.
$$
Let us define the functions $G(\zeta,\eta)$ and $G^{s}(\zeta,\eta)$ by
$$
G(\zeta,\eta)=k\left(\frac{\zeta+\eta}{2},\frac{\zeta-\eta}{2}\right),\qquad G^{s}(\zeta,\eta)=s\left(\frac{\zeta+\eta}{2},\frac{\zeta-\eta}{2}\right)
$$
and denote by $\ds b_{i}(\zeta,\eta)=\rho_{i}\left(\frac{\zeta+\eta}{2},\frac{\zeta-\eta}{2}\right)$ for $i=1,\ldots,5$ and 
$$
g_{1}(\zeta)=m\left(\frac{\zeta}{2}\right)\quad\text{and}\quad g_{2}(\zeta)=m^{s}\left(\frac{\zeta}{2}\right).
$$
From~\eqref{ctw23} and~\eqref{ctw24}, one obtain the partial differential equations
\begin{equation}\label{ctw25}
\left\{\begin{array}{rcl}
G_{\zeta\eta}(\zeta,\eta)&=&b_{1}(G_{\zeta\zeta}^{s}(\zeta,\eta)-2G_{\zeta\eta}^{s}(\zeta,\eta)+G_{\eta\eta}^{s}(\zeta,\eta))+b_{2}G(\zeta,\eta)
\\
&&+b_{3}G^{s}(\zeta,\eta)+b_{4}(G_{\zeta}^{s}(\zeta,\eta)-G_{\eta}^{s}(\zeta,\eta))
\\
G(\zeta,0)&=&g_{1}(\zeta)
\\
G(\zeta,\zeta-2\varepsilon)&=&0,
\end{array}\right.
\end{equation}
and
\begin{equation}\label{ctw26}
\left\{\begin{array}{rcl}
G_{\zeta\eta}^{s}(\zeta,\eta)&=&b_{1}G(\zeta,\eta)+b_{5}G^{s}(\zeta,\eta)
\\
G^{s}(\zeta,0)&=&g_{2}(\zeta)
\\
G^{s}(\zeta,\zeta-2\varepsilon)&=&0.
\end{array}\right.
\end{equation}
Integrating~\eqref{ctw25}, first respect to $\eta$ between $\varepsilon$ and $\eta$, and then with respect to $\zeta$ between $\eta+2\varepsilon$ and $\zeta$, one gets
\begin{equation}\label{ctw27}
\begin{split}
G(\zeta,\eta)=&g_{1}(\zeta)-g_{1}(\eta+2\varepsilon)+\frac{1}{4}\int_{\eta+2\varepsilon}^{\zeta}\int_{0}^{\eta}b_{2}(\tau,s)G(\tau,s)\,\ud s\,\ud\tau
\\
&+\frac{1}{4}\int_{\eta+2\varepsilon}^{\zeta}\int_{0}^{\eta}b_{1}(\tau,s)(G_{\zeta\zeta}^{s}(\tau,s)-2G_{\zeta\eta}^{s}(\tau,s)+G_{\eta\eta}^{s}(\tau,s))\,\ud s\,\ud\tau
\\
&+\frac{1}{4}\int_{\eta+2\varepsilon}^{\zeta}\int_{0}^{\eta}b_{3}(\tau,s)G^{s}(\tau,s)+b_{4}(\tau,s)(G_{\zeta}^{s}(\tau,s)-G_{\eta}^{s}(\tau,s))\,\ud s\,\ud\tau.
\end{split}
\end{equation}
In the same way, we integrate~\eqref{ctw26} first respect to $\eta$ between $\varepsilon$ and $\eta$, and then with respect to $\zeta$ between $\eta$ and $\zeta$, one gets
\begin{equation}\label{ctw28}
G^{s}(\zeta,\eta)=g_{2}(\zeta)-g_{2}(\eta+2\varepsilon)+\int_{\eta+2\varepsilon}^{\zeta}\int_{0}^{\eta}b_{1}(\tau,s)G(\tau,s)+b_{5}(\tau,s)G^{s}(\tau,s))\,\ud s\,\ud\tau.
\end{equation}

We use a classical iterative method in order to prove that the coupled equation~\eqref{ctw27} and~\eqref{ctw28} have a unique solution. Let us define the functions $G^{0}$ and $G^{s,0}$ as
$$
G^{0}(\zeta,\eta)=g_{1}(\zeta)-g_{1}(\eta+2\varepsilon)\qquad\text{and}\qquad G^{s,0}(\zeta,\eta)=g_{2}(\zeta)-g_{2}(\eta+2\varepsilon)
$$
and set up the following recursion for $n=0,1,2,\ldots$
\begin{equation*}
\begin{split}
G^{n+1}(\zeta,\eta)&=\frac{1}{4}\int_{\eta+2\varepsilon}^{\zeta}\int_{0}^{\eta}b_{2}(\tau,s)G^{n}(\tau,s)\,\ud s\,\ud\tau
\\
&+\frac{1}{4}\int_{\eta+2\varepsilon}^{\zeta}\int_{0}^{\eta}b_{1}(\tau,s)(G_{\zeta\zeta}^{s,n}(\tau,s)-2G_{\zeta\eta}^{s,n}(\tau,s)+G_{\eta\eta}^{s,n}(\tau,s))\,\ud s\,\ud\tau
\\
&+\frac{1}{4}\int_{\eta+2\varepsilon}^{\zeta}\int_{0}^{\eta}b_{3}(\tau,s)G^{s,n}(\tau,s)+b_{4}(\tau,s)(G_{\zeta}^{s,n}(\tau,s)-G_{\eta}^{s,n}(\tau,s))\,\ud s\,\ud\tau,
\end{split}
\end{equation*}
and
\begin{equation*}
G^{s,n+1}(\zeta,\eta)=\frac{1}{4}\int_{\eta+2\varepsilon}^{\zeta}\int_{0}^{\eta}b_{1}(\tau,s)G^{n}(\tau,s)+b_{5}(\tau,s)G^{s,n}(\tau,s))\,\ud s\,\ud\tau.
\end{equation*}
By defining $M=\max\{2\|g_{1}'\|_{L^{\infty}},2\|g_{2}'\|_{L^{\infty}},\|g_{2}''\|_{L^{\infty}}\}$, we obtain
\begin{equation*}
\begin{split}
&|G^{0}(\zeta,\eta)|=|g_{1}(\zeta)-g_{1}(\eta)|\leq\|g_{1}'\|_{L^{\infty}}|\zeta-\eta|\leq M,
\\
&|G^{s,0}(\zeta,\eta)|=|g_{2}(\zeta)-g_{2}(\eta)|\leq\|g_{2}'\|_{L^{\infty}}|\zeta-\eta|\leq M,
\\
&|G^{s,0}_{\zeta}(\zeta,\eta)|=|g_{2}'(\zeta)|\leq M,\quad|G^{s,0}_{\eta}(\zeta,\eta)|=|g_{2}'(\zeta)|\leq M,
\\
&|G^{s,0}_{\eta\eta}(\zeta,\eta)|=|g_{2}''(\eta)|\leq M,\quad|G^{s,0}_{\zeta\zeta}(\zeta,\eta)|=|g_{2}''(\zeta)|\leq M,
\\
&|G^{s,0}_{\eta\zeta}(\zeta,\eta)|=0.
\end{split}
\end{equation*}
Let us denote by $\delta=\max\{\xi,2\epsilon\}$ and
$$
K=\frac{1}{2}\max\left\{\|b_{1}\|_{\mathrm{C}^{1}}+\|b_{5}\|_{\mathrm{C}^{1}},4\|b_{1}\|_{L^{\infty}}+\|b_{2}\|_{L^{\infty}}+\|b_{3}\|_{L^{\infty}}+2\|b_{4}\|_{L^{\infty}}\right\},
$$
and suppose that for some $n\in\N$ we have
\begin{equation}\label{ctw29}
\begin{array}{ll}
\ds|G^{n}(\zeta,\eta)|\leq\frac{M.K^{n}}{n!}(\eta+\delta)^{n},&\ds|G^{s,n}(\zeta,\eta)|\leq\frac{M.K^{n}}{n!}(\eta+\delta)^{n},
\\
\ds|G_{\zeta}^{s,n}(\zeta,\eta)|\leq\frac{M.K^{n}}{n!}(\eta+\delta)^{n},&\ds|G_{\eta}^{s,n}(\zeta,\eta)|\leq\frac{M.K^{n}}{n!}(\eta+\delta)^{n},
\\
\ds|G_{\zeta\zeta}^{s,n}(\zeta,\eta)|\leq\frac{M.K^{n}}{(n-1)!}(\eta+\delta)^{n-1},&\ds|G_{\zeta\eta}^{s,n}(\zeta,\eta)|\leq\frac{M.K^{n}}{(n-1)!}(\eta+\delta)^{n-1},
\\
\ds|G_{\eta\eta}^{s,n}(\zeta,\eta)|\leq\frac{M.K^{n}}{(n-1)!}(\eta+\delta)^{n-1}.&
\end{array}
\end{equation}
From~\eqref{ctw27} and~\eqref{ctw28}, we obtain
\begin{equation*}
\begin{split}
|G^{s,n+1}(\zeta,\eta)|&\leq\frac{1}{4}\|b_{1}\|_{L^{\infty}}\int_{\eta+2\varepsilon}^{\zeta}\int_{0}^{\eta}|G^{n}(\tau,s)|\,\ud s\,\ud\tau+\frac{1}{4}\|b_{5}\|_{L^{\infty}}\int_{\eta+2\varepsilon}^{\zeta}\int_{0}^{\eta}|G^{s,n}(\tau,s)|\,\ud s\,\ud\tau
\\
&\leq\left(\frac{\|b_{1}\|_{L^{\infty}}+\|b_{5}\|_{L^{\infty}}}{4}\right)\frac{MK^{n}}{n!}\int_{\eta+2\varepsilon}^{\zeta}\int_{0}^{\eta}(\tau+s)^{n}\,\ud s\,\ud\tau
\\
&\leq\left(\frac{\|b_{1}\|_{L^{\infty}}+\|b_{5}\|_{L^{\infty}}}{4}\right)\frac{MK^{n}}{(n+1)!}\int_{\eta+2\varepsilon}^{\zeta}(\tau+\eta)^{n+1}-\tau^{n+1}\,\ud\tau
\\
&\leq\left(\frac{\|b_{1}\|_{L^{\infty}}+\|b_{5}\|_{L^{\infty}}}{2}\right)\frac{MK^{n}}{(n+1)!}(\eta+\delta)^{n+1},
\end{split}
\end{equation*}
and
\begin{equation*}
\begin{split}
|G^{n+1}(\zeta,\eta)|&\leq\frac{\|b_{2}\|_{L^{\infty}}}{4}\int_{\eta+2\varepsilon}^{\zeta}\int_{0}^{\eta}|G^{n}(\tau,s)|\,\ud s\,\ud\tau+\frac{\|b_{3}\|_{L^{\infty}}}{4}\int_{\eta+2\varepsilon}^{\zeta}\int_{0}^{\eta}|G^{s,n}(\tau,s)|\,\ud s\,\ud\tau
\\
&+\frac{\|b_{1}\|_{L^{\infty}}}{4}\int_{\eta+2\varepsilon}^{\zeta}\int_{0}^{\eta}|G_{\zeta\zeta}^{s,n}(\tau,s)|+2|G_{\zeta\eta}^{s,n}(\tau,s)|+|G_{\eta\eta}^{s,n}(\tau,s)|\,\ud s\,\ud\tau
\\
&+\frac{\|b_{4}\|_{L^{\infty}}}{4}\int_{\eta+2\varepsilon}^{\zeta}\int_{0}^{\eta}|G_{\zeta}^{s,n}(\tau,s)|+|G_{\eta}^{s,n}(\tau,s)|\,\ud s\,\ud\tau
\\
&\leq\left(\frac{\|b_{2}\|_{L^{\infty}}+\|b_{3}\|_{L^{\infty}}+2\|b_{4}\|_{L^{\infty}}}{4}\right)\frac{MK^{n}}{n!}\int_{\eta+2\varepsilon}^{\zeta}\int_{0}^{\eta}(\tau+s)^{n}\,\ud s\,\ud\tau
\\
&+\|b_{1}\|_{L^{\infty}}\frac{MK^{n}}{(n-1)!}\int_{\eta+2\varepsilon}^{\zeta}\int_{0}^{\eta}(\tau+s)^{n-1}\,\ud s\,\ud\tau
\\
&\leq\left(\frac{\|b_{2}\|_{L^{\infty}}+\|b_{3}\|_{L^{\infty}}+2\|b_{4}\|_{L^{\infty}}+4\|b_{1}\|_{L^{\infty}}}{2}\right)\frac{MK^{n}}{(n+1)!}(\eta+\delta)^{n+1}.
\end{split}
\end{equation*}
In a similar way, we obtain
\begin{equation*}
\begin{split}
|G_{\zeta}^{s,n+1}(\zeta,\eta)|&\leq\left(\frac{\|b_{1}\|_{L^{\infty}}+\|b_{5}\|_{L^{\infty}}}{4}\right)\frac{MK^{n}}{(n+1)!}(\eta+\delta)^{n+1},
\\
|G_{\eta}^{s,n+1}(\zeta,\eta)|&\leq\left(\frac{\|b_{1}\|_{L^{\infty}}+\|b_{5}\|_{L^{\infty}}}{4}\right)\frac{MK^{n}}{(n+1)!}(\eta+\delta)^{n+1},
\\
|G_{\zeta\eta}^{s,n+1}(\zeta,\eta)|&\leq\left(\frac{\|b_{1}\|_{L^{\infty}}+\|b_{5}\|_{L^{\infty}}}{4}\right)\frac{MK^{n}}{n!}(\eta+\delta)^{n},
\end{split}
\end{equation*}
$$
|G_{\zeta\zeta}^{s,n+1}(\zeta,\eta)|\leq\left(\frac{\|b_{1}\|_{L^{\infty}}+\|b_{5}\|_{L^{\infty}}+\|b_{1\zeta}\|_{L^{\infty}}+\|b_{5\zeta}\|_{L^{\infty}}}{4}\right)\frac{MK^{n}}{n!}(\eta+\delta)^{n},
$$
and
$$
|G_{\eta\eta}^{s,n+1}(\zeta,\eta)|\leq\left(\frac{\|b_{1}\|_{L^{\infty}}+\|b_{5}\|_{L^{\infty}}+\|b_{1\eta}\|_{L^{\infty}}+\|b_{5\eta}\|_{L^{\infty}}}{4}\right)\frac{MK^{n}}{n!}(\eta+\delta)^{n}.
$$
Thus, by induction we have proved that~\eqref{ctw29} for every $n\in\N$. Once the estimates~\eqref{ctw29} are proved, it follows that the series
$$
G^{s}(\zeta,\eta)=\sum_{n=0}^{\infty}G^{s,n}(\zeta,\eta),\qquad G(\zeta,\eta)=\sum_{n=0}^{\infty}G^{n}(\zeta,\eta),
$$
which converge absolutely and uniformly in $\{(\zeta,\eta),\;0\leq\eta\leq\zeta\leq2\xi\}$ or $\{(\zeta,\eta),\;\xi\leq\eta\leq\zeta\leq 2\}$ and their sum define two continuous functions which are solution of~\eqref{ctw27} and~\eqref{ctw28}. To see that these functions are indeed more regular, we use the equations they satisfy. Indeed, from~\eqref{ctw27}, we see that $G^{s}$ belongs to $\mathrm{C}^{2}$ if $b_{1}$ and $b_{5}$ are continuous. Then, from~\eqref{ctw27}, we see that if $b_{i}$ with $i=1,\ldots,5$ are continuous functions, then $G$ belong to $\mathrm{C}^{2}$. Finally, by the method of successive approximations we can show that the equations~\eqref{ctw27} and~\eqref{ctw28} have a unique continuous solution.
\end{pr}
\section{Exponential stability of the closed-loop system}\label{ctw32}
Let us define the map
\begin{equation*}
\begin{array}{rl}
\Pi:H_{L}^{1}(0,\xi)\!\!\times\!\! H_{R}^{1}(\xi,1)\!\!\times\!\! L^{2}(0,\xi)\!\!\times\!\! L^{2}(\xi,1)\longrightarrow&H_{L}^{1}(0,\xi)\!\!\times\!\! H_{R}^{1}(\xi,1)\!\!\times\!\! L^{2}(0,\xi)\!\!\times\!\! L^{2}(\xi,1)
\\
(u_{1},u_{2},v_{1},v_{2})\longmapsto&\Pi(u_{1},u_{2},v_{1},v_{2})=(w_{1},w_{2},z_{1},z_{2})
\end{array}
\end{equation*}
where
\begin{equation*}
\begin{split}
w_{1}(x)&=h_{1}(x)u_{1}(x)-\int_{0}^{x}k_{1}(x,y)u_{1}(y)\,\ud y-\int_{0}^{x}s_{1}(x,y)v_{1}(y)\,\ud y,
\\
w_{2}(x)&=h_{2}(x)u_{2}(x)-\int_{1}^{x}k_{2}(x,y)u_{2}(y)\,\ud y-\int_{1}^{x}s_{2}(x,y)v_{2}(y)\,\ud y,
\\
z_{1}(x)&=s_{1y}(x,x)u_{1}(x)-s_{1}(x,x)u_{1}'(x)+h_{1}(x)v_{1}(x)
\\
&-\int_{0}^{x}[2\lambda_{1}(y)s_{1}(x,y)+k_{1}(x,y)]v_{1}(y)\,\ud y-\int_{0}^{x}[\beta_{1}(y)s_{1}(x,y)+s_{1yy}(x,y)]u_{1}(y)\,\ud y,
\\
z_{2}(x)&=s_{2y}(x,x)u_{2}(x)-s_{2}(x,x)u_{2}'(x)+h_{2}(x)v_{2}(x)
\\
&-\int_{0}^{x}[2\lambda_{2}(y)s_{2}(x,y)+k_{2}(x,y)]v_{2}(y)\,\ud y-\int_{0}^{x}[\beta_{2}(y)s_{2}(x,y)+s_{2yy}(x,y)]u_{2}(y)\,\ud y.
\end{split}
\end{equation*}
This linear map is bounded, and hence there exists a positive constant $C_{1}$ such that
\begin{equation}\label{ctw45}
\begin{split}
\|\Pi(u_{1},u_{2},v_{1},v_{2})\|_{H^{1}(0,\xi)\times H^{1}(\xi,1)\times L^{2}(0,\xi)\times L^{2}(0,\xi)}
\\
\leq C_{1}\|(u_{1},u_{2},v_{1},v_{2})\|_{H^{1}(0,\xi)\times H^{1}(\xi,1)\times L^{2}(0,\xi)\times L^{2}(0,\xi)}.
\end{split}
\end{equation}
The importance of $\Pi$ is that it maps solutions $(u_{1},u_{2},\dot{u}_{1},\dot{u}_{2})$ of~\eqref{ctw2} with the feedback $U(t)$ given by~\eqref{ctw5} into solution $(w_{1},w_{2},\dot{w}_{1},\dot{w}_{2})$ of~\eqref{ctw6}. The map $\Pi$ which converting the original unstable system into the target system, is invertible. Indeed, to obtain the kernel functions $\hat{k}_{1}(x,y)$, $\hat{k}_{2}(x,y)$, $\hat{s}_{1}(x,y)$ and $\hat{s}_{2}(x,y)$ defining $\Pi^{-1}$, we simply replace the functions $d_{1}(x)$ by $-\lambda_{1}(x)$, $d_{2}(x)$ by $-\lambda_{2}(x)$, $\lambda_{1}(x)$ by $-d_{1}(x)$, $\lambda_{2}(x)$ by $-d_{2}(x)$ and changing the role of $c_{1}$, and $\beta_{1}$ and $c_{2}$ and $\beta_{2}$ in the previous analysis for the kernels $k_{1}(x, y)$, $k_{2}(x,y)$, $s_{1}(x, y)$ and $s_{2}(x, y)$. Thus, we get a map
\begin{equation*}
\begin{array}{rl}
\Pi^{-1}:H_{L}^{1}(0,\xi)\!\!\times\!\! H_{R}^{1}(\xi,1)\!\!\times\!\! L^{2}(0,\xi)\!\!\times\!\! L^{2}(\xi,1)\longrightarrow&H_{L}^{1}(0,\xi)\!\!\times\!\! H_{R}^{1}(\xi,1)\!\!\times\!\! L^{2}(0,\xi)\!\!\times\!\! L^{2}(\xi,1)
\\
(w_{1},w_{2},z_{1},z_{2})\longmapsto&\Pi^{-1}(w_{1},w_{2},z_{1},z_{2})=(u_{1},u_{2},v_{1},v_{2})
\end{array}
\end{equation*}
and a positive constant $C_{2}$ such that
\begin{equation}\label{ctw46}
\begin{split}
\|\Pi^{-1}(w_{1},w_{2},z_{1},z_{2})\|_{H^{1}(0,\xi)\times H^{1}(\xi,1)\times L^{2}(0,\xi)\times L^{2}(0,\xi)}
\\
\leq C_{2}\|(w_{1},w_{2},z_{1},z_{2})\|_{H^{1}(0,\xi)\times H^{1}(\xi,1)\times L^{2}(0,\xi)\times L^{2}(0,\xi)}.
\end{split}
\end{equation}

It is well known that if the initial data $(w_{1}^{0},w_{2}^{0},w_{1}^{1},w_{2}^{2})$ of the system~\eqref{ctw6} belongs to the domain
\begin{equation*}
\begin{split}
D=\{(w_{1},w_{2},z_{1},z_{2})\in H_{L}^{2}(0,\xi)\times H_{R}^{2}(\xi,1)\times H_{L}^{1}(0,\xi)\times H_{R}^{1}(\xi,1):w_{1}(\xi)=w_{2}(\xi),
\\
w_{1}'(\xi)=w_{2}'(\xi)\}
\end{split}
\end{equation*}
then $w(x,t)=\mathbb{1}_{(0,\xi)}w_{1}(x,t)+\mathbb{1}_{(\xi,1)}w_{2}(x,t)$ belongs to $H_{0}^{1}(0,1)\cap H^{2}(0,1)$ and satisfying the following boundary problem
\begin{equation}\label{ctw44}
\left\{\begin{array}{ll}
\ddot{w}(x,t)=w_{1}''(x,t)-2d(x)\dot{w}(x,t)-c(x)w(x,t)&\text{in }(0,1)\times(0,+\infty)
\\
w(0,t)=w(1,t)=0&\text{for } t\in (0,+\infty)
\\
w(x,0)=w^{0}(x),\;\dot{w}(x,0)=w^{1}(x)&\text{for }x\in(0,1),
\end{array}\right.
\end{equation}
where the initial data are given by $w^{0}(x)=\mathbb{1}_{(0,\xi)}w_{1}^{0}(x)+\mathbb{1}_{(\xi,1)}w_{2}^{0}(x)$ belongs to $H_{0}^{1}(0,1)\cap H^{2}(0,1)$ and $w^{1}(x)=\mathbb{1}_{(0,\xi)}w_{1}^{1}(x)+\mathbb{1}_{(\xi,1)}w_{2}^{1}(x)$ belongs to $H_{0}^{1}(0,1)$ and the coefficients are given by $d(x)=\mathbb{1}_{(0,\xi)}d_{1}(x)+\mathbb{1}_{(\xi,1)}d_{2}(x)$ and $c(x)=\mathbb{1}_{(0,\xi)}c_{1}(x)+\mathbb{1}_{(\xi,1)}c_{2}(x)$. Moreover, Since $\|(w_{1}(\,.\,,t),w_{2}(\,.\,,t),\dot{w}_{1}(\,.\,,t),\dot{w}_{2}(\,.\,,t))\|_{H^{1}(0,\xi)\times H^{1}(\xi,1)\times L^{2}(0,\xi)\times L^{2}(\xi,1)}$ is equal to $\|(w(\,.\,,t),\dot{w}(\,.\,,t))\|_{H^{1}(0,1)\times L^{2}(0,1)}$ for every $t\geq 0$ and the solution $(w,\dot{w})$ of the system~\eqref{ctw44} is exponentially stable, i.e. there exist two constants $C>0$ and $\omega>0$ such that for every $(w^{0},w^{1})\in H^{1}(0,1)\times L^{2}(0,1)$ we have
\begin{equation}\label{ctw50}
\|(w(\,.\,,t),\dot{w}(\,.\,,t))\|_{H^{1}(0,1)\times L^{2}(0,1)}\leq C\e^{-\omega t}\|(w^{0},w^{1})\|_{H^{1}(0,1)\times L^{2}(0,1)},
\end{equation}
then for every $(w_{1}^{0},w_{1}^{0},w_{1}^{1},w_{2}^{1})\in D$ we have
\begin{equation}\label{ctw47}
\begin{split}
\|(w_{1}(\,.\,,t),w_{2}(\,.\,,t),\dot{w}_{1}(\,.\,,t),\dot{w}_{2}(\,.\,,t))\|_{H^{1}(0,\xi)\times H^{1}(\xi,1)\times L^{2}(0,\xi)\times L^{2}(\xi,1)}
\\
\leq C\e^{-\omega t}\|(w_{1}^{0},w_{2}^{0},w_{0}^{1},w_{2}^{1})\|_{H^{1}(0,\xi)\times H^{1}(\xi,1)\times L^{2}(0,\xi)\times L^{2}(\xi,1)},
\end{split}
\end{equation}
where by density of the domain $D$ in the energy space $H^{1}(0,\xi)\times H^{1}(\xi,1)\times L^{2}(0,\xi)\times L^{2}(\xi,1)$ the estimate remain valid for every $(w_{1}^{0},w_{1}^{0},w_{1}^{1},w_{2}^{1})\in H^{1}(0,\xi)\times H^{1}(\xi,1)\times L^{2}(0,\xi)\times L^{2}(\xi,1)$. The functions $d_{1},\,d_{2},\,c_{1}$ and $c_{2}$ are part of the design of the feedback law, and hence we are able to consider~\eqref{ctw44} with constant coefficients. In this case, for any $\omega>0$, we can find the parameters $d_{1},\,d_{2},\,c_{1}$ and $c_{2}$ so that~\eqref{ctw7} holds. Indeed, the constant coefficients case, where we can choose the parameters $d_{1}=d_{2}=d$ and $c_{1}=c_{2}=c$ so that the exponential decay rate $\omega$ in~\eqref{ctw50} is as large as desired (in particular $\omega=d$ see~\cite{SCK,CZ}), thus combining~\eqref{ctw45},~\eqref{ctw46} and~\eqref{ctw47}, estimate~\eqref{ctw7} holds easily with the compatibility conditions~\eqref{ctw48} and~\eqref{ctw49}.

Finally, problem~\eqref{ctw2} with the feedback defined in~\eqref{ctw5} is well posed since it can be transformed to the problem~\eqref{ctw6} via the isomorphism $\Pi$ defined above, where the problem~\eqref{ctw6} is well posed and its solution is written in term of the semigroup. Hence, the regularity of the solution given by Theorem~\ref{ctw51} holds too. 

\end{document}